\documentclass[10pt,a4paper]{article}

\usepackage{indentfirst}			
 \usepackage{enumerate}				
\usepackage{amsmath, amsgen,amsfonts,amssymb,amsbsy}	

\def\PRD{\it Phys.\ Rev.\ D\ }

\begin{document}

\thispagestyle{plain}		

\title{Harada-Maxwell Static Spherically Symmetric Spacetimes}
\author{Alan Barnes\\ 
26 Havannah Lane, \\
Congleton CW12 2EA, \\
United Kingdom. \\
E-Mail: \ \ {\tt Alan.Barnes45678{\bf @}gmail.com}}
\maketitle

\begin{abstract}
\noindent Very recently Harada has proposed a gravitational theory which is of third
order in the derivatives of the metric tensor with the property that any solution
of Einstein's field equations (EFEs) possibly with a cosmological constant
is necessarily a solution of the new theory. Remarkably he showed that even in a matter-dominated universe with zero
cosmological constant, there is a late-time transition from decelerating to
accelerating expansion.

\noindent Harada also derived an exact solution which is generalisation of the
Schwarzschild solution. However, this was not the most general static spherically
vacuum solution of the theory and the general solution was subsequently obtained by
Barnes.  

\noindent Recently Tarciso et al.\ have considered regular black holes in Harada's
theory coupled to non-linear electrodynamics and scalar fields. In particular they
exhibit a four-parameter solution with a zero scalar field whose source is a Maxwell
electromagnetic field. It is a straightforward generalisation of Harada's vacuum
solution analagous to the Reissner-Nordstr\"om generalistaion of the Schwarzschild
solution.

\noindent However, this solution is not the most general static spherically symmetric
solution of the Harada-Maxwell field equations (i.e.\ Harada gravitational fields with
a Maxwell electromagnetic source). The most general such solution is obtained in this
paper.
\end{abstract}

\vspace{-5 pt}
\section{Introduction}
\noindent Recently Harada\cite{harada} has proposed a new gravitational theory
which involves the totally symmetric derivatives
of, firstly a trace-modified  Einstein tensor $\tilde{G}_{ab}$:
\begin{equation}
  H_{abc} = \tilde{G}_{(ab;c)}\quad\mathrm{where} \quad
  \tilde{G}_{ab} = R_{ab}-\frac{1}{3}Rg_{ab}=   G_{ab}-\frac{1}{6}Gg_{ab}
  \label{tildeG}
\end{equation}
where round brackets indicate symmetrisation; and secondly, the similarly modified
energy-momentum tensor:
\begin{equation}
  T_{abc} = \tilde{T}_{(ab;c)} \quad\mathrm{where} \quad
  \tilde{T}_{ab} = T_{ab}-\frac{1}{6}Tg_{ab}.
  \label{tildeT}
\end{equation}
As the field equations of the theory he proposed:
\begin{equation}
  H_{abc} = T_{abc}.
  \label{hfe}
\end{equation}

The energy-momentum conservation equation follows from \eqref{hfe} by contraction:
\[ g^{ac}H_{abc}=G^a_{b;a} = 0 = g^{ac}T_{abc} =T^a_{b;a}. \]
It also follows immediately  that any solution of the EFEs with or without a cosmological
constant automatically satisfies equation \eqref{hfe}. 

Remarkably Harada\cite{harada}\cite{harad2} showed that even in a matter-dominated
universe with zero cosmological constant, there is a late-time transition from
decelerating to accelerating expansion.

Mantica and Molinari\cite{mantmol} have examined Harada's field equations and
shown that they can be recast in the form of Einstein's field equations with an
additional matter source term which is a second-order conformal Killing tensor.  
Consequently they named the theory `Conformal Killing Gravity'.

The  vacuum case in the theory is characterised by the condition $T_{abc} =0$.
In this paper solutions where the source is an electromagnetic field satisfying
Maxwell's equations are considered.  By analogy with Einstein-Maxwell fields these will be
referred to below as `Harada-Maxwell fields'.

\section{Static Spherically Symmetric Harada-Maxwell Solutions}
\noindent Among the metrics discussed by Tarciso et al.\cite{tarciso}, they exhibit
\begin{equation}
  \mathrm{d}s^2 = e^{2a(r)}\mathrm{d}t^2 - e^{-2a(r)}\mathrm{d}r^2 -
      r^2(\mathrm{d}\theta^2+\sin^2\theta\mathrm{d}\phi^2),
   \label{sssmet}
\end{equation}   
where \begin{equation}
  e^{2a(r)} = 1-2m/r+q/r^2-\Lambda r^2/3-\lambda r^4/5.
  \label{rnlike}
\end{equation}
 $m$, $q$, $\Lambda$ and $\lambda$ are arbitrary constants of integration. When
$\lambda =0$, this is the well-known Reissner-Nordstr\"om metric generalised to include
a cosmological constant term, but the $\lambda r^4/5$ term is a new feature of Harada's
theory. Tarciso et al.\ use what may be described as a variant of Synge's g-method\cite
{synge} in that they assume a form of the metric and then use the field equations to
calculate the form of the associated energy-momentum tensor and Lagrangian.

In this paper a more direct approach is used: namely Synge's T-method\cite{synge}. The Harada
field equations with a Maxwell field as source are then solved for the static spherically
 symmetric case.  In terms of curvature coordinates the most general spherically 
symmetric static metric is
\begin{equation}
  \mathrm{d}s^2 = e^{2a}\mathrm{d}t^2 - e^{2b}\mathrm{d}r^2 -
      r^2(\mathrm{d}\theta^2+\sin^2\theta\mathrm{d}\phi^2),
   \label{gensssmet}
\end{equation}
where $a$ and $b$ are functions of $r$ only. Subsequent calculations will use the
\emph{frame} components of tensors in the obvious Lorentz orthonormal tetrad of one forms:
\begin{equation}
  e^a\mathrm{d}t,\qquad e^b\mathrm{d}r,\qquad r\mathrm{d}\theta, \qquad
  r\sin\theta\mathrm{d}\phi
\label{tetrad}
\end{equation}
It will be assumed that the Maxwell field is also spherically symmetric, i.e.\ invariant
under  SO(3), and that there are no magnetic monopoles so that $F_{23}=0$.
Thus the only non-zero components the antisymmetric field tensor $F_{ab}$ are\\
$F_{01} = -F_{10} = f(t,r)$ where $f(t,r)$ is an arbitrary function of $t$ and $r$.
Maxwell's equations $F^{ab}_{\ \ ;b}=0$ and $F_{[ab;c]}=0$ then imply that $F_{01} = \sqrt{8\pi}q/r^2$,
where $q$ is a constant; the factor $\sqrt{8\pi}$ being introduced for later convenience.
Note that it was not necessary to assume the time independence of the electromagnetic
field as this follows from Maxwell's equations and the assumption of spherical symmetry.

In fact, the assumption that there are no magnetic monopoles is not essential as shown in
the textbook by Pleba\'nski \& Krasi\'nski\cite{Pleb}; a duality rotation may used to
produce a new field $\tilde F_{ab}$ for which $\tilde F_{\theta\phi}=0$ and
$\tilde q = \sqrt{q^2+k^2}$ where the constant $k$ is the magnetic monopole strength.
The electromagnetic energy-momentum tensor is not changed by the duality rotation. 

The energy-momentum tensor is given by
\begin{equation}
  T_{00} = -T_{11} = T_{22}= T_{33} = q^2/r^4
\end{equation}
and it may be verified that $T=0$ and $T^{ab}_{\ \ ;b}=0$ as expected.
As the tensor $T_{abc}$ is totally symmetric, essentially its only non-zero \emph{frame}
components are
\begin{equation}
  T_{111} = 12q^2e^{-b}/r^5,\qquad T_{100} =-T_{111}/3, \qquad
  T_{122}=T_{133}=-2T_{111}/3.
\end{equation}
Similarly the only non-zero \emph{frame} components of $H_{abc}$ are $H_{100}$,
$H_{111}$ and $H_{122}=H_{133}$. In fact only two of these are
linearly independent as $2H_{122}-H_{100}+H_{111} \equiv 0$.
It may be noticed in passing that the above equations for the components of $H_{abc}$ are
valid for any \emph{static spherically symmetric metric in curvature coordinates} and so
it may be possible to generalise the analysis below for more exotic matter fields.

For Harada-Maxwell fields it is convenient to work with the two Harada field equations:
\begin{equation}
  3H_{100}+H_{111} = r(a''-2a'^2-4a'b'+b''-2b'^2)-a'-b' =0
  \label{eqn1}
\end{equation}
and
\begin{eqnarray}
  H_{100}-T_{100} &=& -r^3(a'''+2a''a'-3a''b'-2a'^2b'-a'b''+2a'b'^2)\nonumber\\ 
  & & + r^2(4a''-8a'b'+2b''-4b'^2)-r(4a'+6b')\nonumber\\
  & & + 4e^{2b} + 4q^2e^{-b}/r^5 -4 = 0,
  \label{eqn2}
\end{eqnarray}
where a prime denotes differentiation with wrt $r$.
The computer algebra system Classi (see \cite{Aman} and \cite{ACGR}) was used to
calculate the components of the tensor $H_{abc}$ and $T_{abc}$ and the subsequent
two field equations.

Using the substitution $b=f-a$ and cancelling common factors, equation\eqref{eqn1}
simplifies to
\begin{equation}
  r f''-2r f'^2 -f' =0.
  \label{feqn}
\end{equation}
Hence
\begin{equation}
  f=-log(c+dr^2)/2
  \label{genf}
\end{equation}
where $c$ and $d$ are arbitrary constants of integration.
Hence the metric takes the form \eqref{gensssmet} with
$e^{2b} = e^{-2a}/(c+dr^2)$.  Eliminating $b$ from equation\eqref{eqn2}, applying the
substitution $y=e^{2a}$, and removing non-zero factors results in the linear equation:
\begin{equation}
  (c+dr^2)r^3y''' - (2c-dr^2)r^2y''-(2c+dr^2)ry'+ 8cy = 8-24q^2/r^2
  \label{gensol}
\end{equation}
If $c \neq 0$, after a suitable constant rescaling of the
$t$ coordinate, $c$ may be set to $\pm $1 without loss of generality.
Similarly if $c=0$, rescaling of the $t$ coordinate may be used to set
$d=1$. Note setting $d = -1$ results in  non-Lorentzian signature of$g_{ab}$).
Thus only one of the arbitrary constants $c$ and $d$ is essential.

When $d=0$ and setting $c=1$ \eqref{gensol} may be reduced to
\begin{equation}
  r^3y'''-2r^2y''-2ry'+8y = 8-24q^2/r^2.
\end{equation}
Noting that the LHS of this equation is homogeneous, one may easily obtain the solution
$y= e^{2a}=e^{-2b}= 1-2m/r+q^2/r^2-\Lambda r^2/3-\lambda r^4/5$
where $m$, $\Lambda$ and $\lambda$ are arbitrary constants.
Again the conditions $c=-1$ and $d=0$ cannot occur for a Lorentzian metric.
The numerical factors of the $1/r$, $r^2$ and $r^4$ terms have been chosen to
correspond with those of Harada's vacuum solution solution.
This is the  first solution exhibited by Tarciso et al.\cite{tarciso}
and bears the same relation to Harada's vacuum solution as does the
Reissner-Nordstr\"om solution to the Schwarzschild solution.

When $c=0$ and setting $d=1$ \eqref{gensol} now reduces to
\begin{equation}
  r^3y'''+r^2y''-ry' = 8/r^2-24q^2/r^4.
\end{equation}
Homogeneity of the LHS again means this  may be easily solved to yield
\begin{equation}
  y =e^{2a} = \lambda -\Lambda r^2/3 + m \log r -1/(2r^2)+q^2/(4r^4)\qquad
  e^{2b} = e^{-2a}/r^2
  \label{logsol}
\end{equation}
where $m$, $\Lambda$ and $\lambda$ are again arbitrary constants.

In the general case, again setting $c=1$, a \emph{particular integral} of \eqref{gensol}
is easily seen to be $y=1+(1+2dr^2)q^2/r^2$.
The \emph{complementary function} is obtained by the Frobenius method (see, for example,
\cite{stephenson}). A solution of the form
\begin{equation}
  y=\sum_{n=0}^\infty a_nr^{\alpha+n}
\end{equation}
is assumed and the analysis here is identical to that in the vacuum case\cite{barnes}.

The \emph{indicial equation} is
\begin{equation}
  (\alpha-2)(\alpha-4)(\alpha+1) =0.
\end{equation}
The case $\alpha=2$ leads to the monomial solution $y=r^2$ (the cosmological
constant term). The cases $\alpha=4$ and $\alpha=-1$ each result in an infinite power
series.  For $n=1$ one obtains $(\alpha^3-2\alpha^2-\alpha+6)a_1=0$ and thus $a_1=0$ in
each case. For $n>=2$ and $\alpha=-1$ the recurrence relation $a_n= -d\frac{n-3}{n}a_{n-2}$
is obtained whilst for $\alpha=4$ one obtains
$a_n=-d\frac{n+2}{n+5}a_{n-2}$.  Clearly in both cases the coefficient $a_n$
vanishes when $n$ is odd and the radius of convergence of each series is
clearly $1/\sqrt{|d|}$.

The general solution of \eqref{gensol} is therefore
\begin{eqnarray}
  y = e^{2a} &=& 1+(1+2dr^2)q^2/r^2 -2mp_1(r)/r +\lambda p_2(r)r^4/5 -\Lambda r^2/3\\
  \mathrm{where} \quad p_1(r)&=& 1 +dr^2/2 -d^2r^4/8+d^3r^6/16\nonumber\\
  & & -5d^4r^8/128+7d^5r^{10}/256\ldots\\
  \mathrm{and}\quad p_2(r) &=& 1 -4dr^2/7 + 8d^2r^4/21-64d^3r^6/231\nonumber\\
  & & +640d^4r^8/3003-512d^5r^{10}/3003  \ldots \\
  \mathrm{with} \quad e^{2b} &=& e^{-2a}/(1+dr^2).
\end{eqnarray}
where $m$, $\Lambda$ and $\lambda$ are again arbitrary constants of integration; the
numerical factors being chosen to correspond with those in Harada's Schwarzschild-like
vacuum solution\cite{harada}.

In an earlier version of this paper and in \cite{barnes}, the possibility of $c=-1$ and $d>0$ was
overlooked. However, this does not in general lead to valid power series solutions. Clearly to maintain Lorentzian
signature $r>1/\sqrt{d}$, but the two power series obtained still have a radius of convergence of
$1/\sqrt{d}$. In fact for $\alpha=-1$ the recurrence relation becomes $a_n= d\frac{n-3}{n}a_{n-2}$
whilst for $\alpha=4$ it is $a_n=d\frac{n+2}{n+5}a_{n-2}$.  The only exception to this is if the constants
$m$ and $\lambda$ are both zero so that the power series terms disappear. This leads to the three parameter
solution
\begin{equation}
  y = e^{2a} = -1+(2dr^2-1)q^2/r^2 -\Lambda r^2/3 \quad \mathrm{with\ \ } e^{2b}= e^{-2a}/(dr^2-1).
\end{equation}

\section{Conclusions}
\noindent All static spherically symmetric Harada-Maxwell fields are derived.
The most general solution involves five essential parameters, three of which may be
identified as the mass $m$ and charge $q$ of the central source plus the cosmological
constant $\Lambda$. The fourth parameter $\lambda$ also appears in Harada's vacuum
solution\cite{harada} whilst the fifth $d$ also appears in the general static spherically
symmetric vacuum solution of Barnes\cite{barnes}. These two parameters have
no analogue in General Relativity. The general solution involves two infinite power
series in the curvature coordinate $r$ which multiply the $m$ and $\lambda$ terms as in
the vacuum case. In fact the only difference between the electromagnetic and vacuum
solutions is an extra term $(1+2dr^2)q^2/r^2$ appearing in $g_{tt}$ and $g_{rr}$. 

The four-parameter `Reissner-Nordstr\"om-like' solution obtained by Tarciso et al. 
\cite{tarciso} corresponds to the case where the fifth parameter $d=0$.  It is a
straightforward electromagnetc generalisation of Harada's `Schwarzschild-like' solution.
In this case the two power series of the general solution degenerate to monomials.

As in the vacuum case a second four-parameter solution may also be obtained which does
not involve power series. This corresponds to the case $c=0,\ d=1$ in \eqref{gensol}. It has logarithmic $r$-dependence and its physical interpretation is currently unclear.

\section*{Acknowledgements}
\noindent The extensive calculations in section 2 were performed using  the
Sheep/Classi package for General Relativity which was kindly supplied to me by Jan
 {\AA}man of the University of Stockholm.  Some calculations were
also performed using the Reduce computer algebra freely available for download
from SourceForge\\(sourceforge.net/projects/reduce-algebra/files/).

\end{document}